\documentclass[conference]{IEEEtran}
\IEEEoverridecommandlockouts
\usepackage{cite}
\usepackage{amsmath,amssymb,amsfonts}
\usepackage{algorithmic}
\usepackage{graphicx}
\usepackage{textcomp}
\usepackage{xcolor}
\usepackage{verbatim}
\usepackage[a4paper, total={184mm,239mm}]{geometry}
\def\BibTeX{{\rm B\kern-.05em{\sc i\kern-.025em b}\kern-.08em
    T\kern-.1667em\lower.7ex\hbox{E}\kern-.125emX}}

\usepackage[colorlinks, linkcolor=red, anchorcolor=blue, citecolor=green]{hyperref}
\usepackage[caption=false]{subfig}
\usepackage{enumitem}
    
\usepackage{tikz}
\usepackage{adjustbox}
\usepackage{array}
\newcommand{\circlednum}[1]{%
  \adjustbox{valign=m}{%
    \tikz[baseline=(char.base)]{
      \node[shape=circle,draw=black,fill=white,text=black,inner sep=0.5pt] (char) {\textbf{#1}};
    }%
  }%
}

\usepackage{booktabs}
\usepackage{makecell}
\usepackage{cellspace}

\usepackage[skip=4pt, size=footnotesize]{caption}
\begin{document}

\title{DARE: An Irregularity-Tolerant Matrix Processing Unit with a \underline{D}ensifying IS\underline{A} and Filtered \underline{R}unahead \underline{E}xecution}


\author{\IEEEauthorblockN{Xin Yang}
\IEEEauthorblockA{\textit{Fudan University} \\
Shanghai, China \\
yangx23@m.fudan.edu.cn}
\and
\IEEEauthorblockN{Xin Fan}
\IEEEauthorblockA{\textit{HKUST} \\
Hong Kong, China \\
xfanat@connect.ust.hk}
\and
\IEEEauthorblockN{Zengshi Wang}
\IEEEauthorblockA{\textit{Fudan University} \\
Shanghai, China \\
wangzs23@m.fudan.edu.cn}
\and
\IEEEauthorblockN{Jun Han}
\IEEEauthorblockA{\textit{Fudan University} \\
Shanghai, China \\
junhan@fudan.edu.cn}
}

\maketitle

\begin{abstract}
Deep Neural Networks (DNNs) are widely applied across domains and have shown strong effectiveness. As DNN workloads increasingly run on CPUs, dedicated Matrix Processing Units (MPUs) and Matrix Instruction Set Architectures (ISAs) have been introduced. At the same time, sparsity techniques are widely adopted in algorithms to reduce computational cost.

Despite these advances, insufficient hardware–algorithm co-optimization leads to suboptimal performance. On the memory side, sparse DNNs incur irregular access patterns that cause high cache miss rates. While runahead execution is a promising prefetching technique, its direct application to MPUs is often ineffective due to significant prefetch redundancy. On the compute side, stride constraints in current Matrix ISAs prevent the densification of multiple logically related sparse operations, resulting in poor utilization of MPU processing elements.

To address these irregularities, we propose DARE, an irregularity-tolerant MPU with a \underline{D}ensifying IS\underline{A} and filtered \underline{R}unahead \underline{E}xecution. DARE extends the ISA to support densifying sparse operations and equips a lightweight runahead mechanism with filtering capability. Experimental results show that DARE improves performance by 1.04$\times$ to 4.44$\times$ and increases energy efficiency by 1.00$\times$ to 22.8$\times$ over the baseline, with 3.91$\times$ lower hardware overhead than NVR.
\end{abstract}


\section{Introduction}

Deep neural networks (DNNs) are widely applied across domains and have shown strong effectiveness. Among their various operations, general matrix multiplication (GEMM) is one of the most compute- and energy-intensive. To reduce the computational cost of GEMM, extensive efforts target both hardware and algorithm design. On the hardware side, as DNN workloads increasingly execute on CPUs~\cite{b_intro1,b_amx_llm,b_lia}, vendors such as Intel and Arm have introduced dedicated matrix Instruction Set Architectures (ISAs), including Advanced Matrix Extensions (AMX)~\cite{b_amx} and Scalable Matrix Extensions (SME)~\cite{b_sme}, along with specialized matrix processing units (MPUs)~\cite{b_rasa,b_vegeta,b_zipper}. On the algorithm side, sparsity has been widely explored, leading to GEMM variants such as sparse matrix–dense matrix multiplication (SpMM) and sampled dense matrix-dense matrix multiplication (SDDMM).

\begin{figure}
    \centering
    \subfloat{\includegraphics[width=1.0\linewidth]{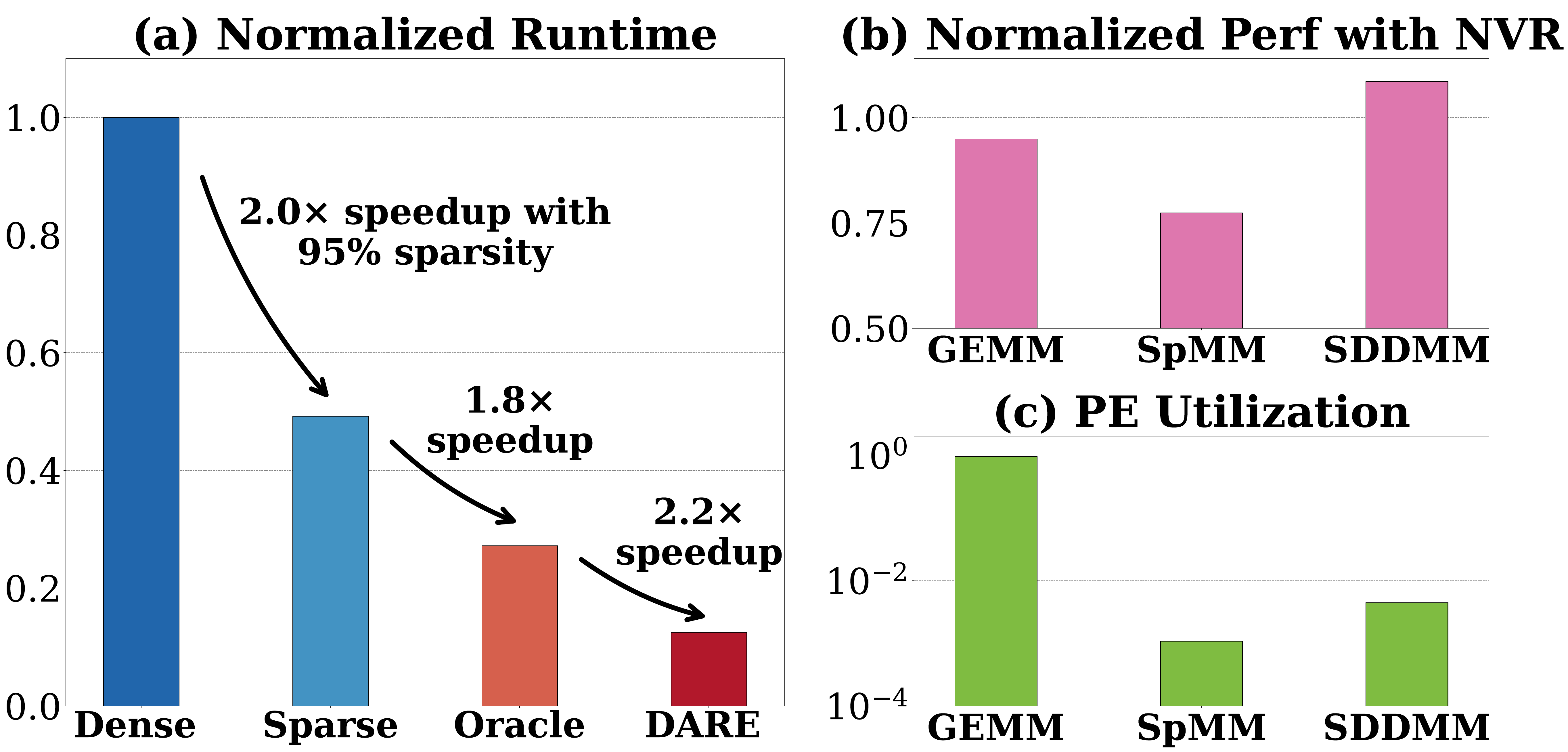}}
    \caption{\textbf{(a)} Runtime of sparse SDDMM normalized to that of dense GEMM on an AMX-like MPU. Oracle assumes a cache without misses. \textbf{(b)} Performance of an MPU with NVR~\cite{b_nvr}, normalized to a baseline MPU without NVR. \textbf{(c)} Processing Element (PE) utilization in a systolic array under various workloads, defined as the ratio of active PEs to the total number of PEs during execution.}
    \label{fig_intro}
\end{figure}

However, insufficient hardware–algorithm co-optimization often results in suboptimal performance. Fig.~\ref{fig_intro}(a) shows that even with 95\% sparsity, SDDMM achieves only a 2.0$\times$ speedup over dense GEMM on an AMX-like MPU. This gap arises from two issues: memory inefficiency and computation irregularity. On the memory side, sparse DNNs typically incur indirect accesses, leading to high cache miss rates and memory stalls. With an ideal zero-miss cache (Oracle in Fig.~\ref{fig_intro}(a)), performance could improve by up to 1.8$\times$. On the compute side, sparse DNNs often employ unstructured sparsity to preserve accuracy, while hardware engines such as systolic arrays, composed of mesh-connected processing elements (PEs), prefer regular and structured sparsity. This mismatch leads to poor hardware utilization and limited acceleration.

For memory inefficiency, runahead execution~\cite{b_rh,b_pre,b_vr,b_dvr,b_svr} is a well-established CPU prefetching technique that mitigates irregular memory accesses by pre-executing stalled instructions. NVR~\cite{b_nvr} first adopted this idea in Neural Processing Units (NPUs) to improve sparse DNN performance. However, Fig.~\ref{fig_intro}(b) shows that directly applying it to Matrix Processing Units (MPUs) may result in performance degradation for workloads such as GEMM and SpMM, compared to an MPU without NVR. In addition, NVR requires 9.72~KB of hardware state, which is a nontrivial overhead for MPUs.


\begin{figure*}
\centerline{\includegraphics[width=1.0\linewidth]{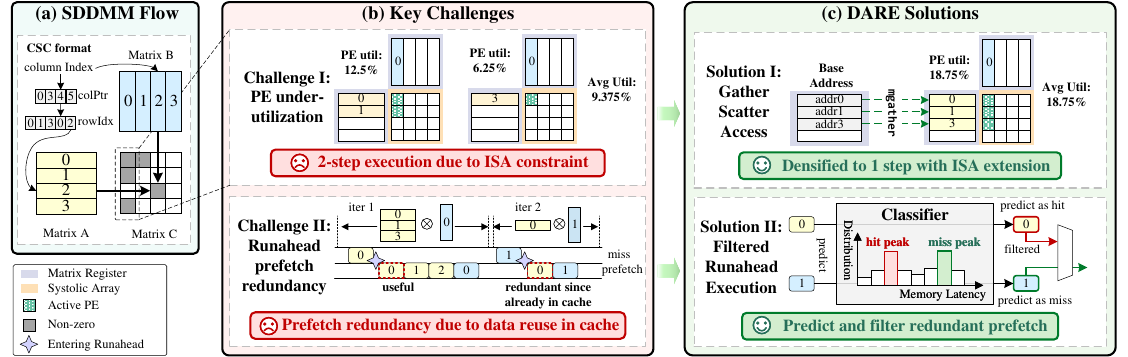}}
\caption{\textbf{(a)} {An exmaple computation flow of SDDMM.} Only the non-zero positions require computation. \textbf{(b)} {The challenges encountered by MPUs with runahead technique on sparse DNN workloads:} PE under-utilization (Section~\ref{sec:isaConstrain}) and runahead prefetch redundancy (Section~\ref{sec:re}). \textbf{(c)} {DARE's solutions to the challenges:} ISA extension to support non-strided access (Section~\ref{sec_isa}) and runahead execution with a runahead filter (Section~\ref{sec_arch}). }
\label{fig_moti}
\end{figure*}

For computation irregularity, Fig.~\ref{fig_intro}(c) shows that sparse workloads often cause low Processing Element (PE) utilization on systolic arrays. Block-wise sparsity \cite{b_block_sparse1,b_block_sparse2,yuan2025native} can improve utilization but may introduce redundant computations. We observe that multiple sparse operations can be logically densified into a single dense operation to better exploit PE resources. However, limitations of current matrix ISAs make such densification infeasible, which further restricts PE utilization.



To address these inefficiencies, we propose DARE, an irregularity-tolerant MPU that integrates a \underline{D}ensifying IS\underline{A} and filtered \underline{R}unahead \underline{E}xecution. Our key contributions are as follows.

\begin{itemize}
    \item We analyze the execution of sparse DNN layers on an MPU with NVR enabled and identify two issues (Fig.~\ref{fig_moti}(b)): (\textbf{i}) PE under-utilization arising from ISA constraints and (\textbf{ii}) redundant NVR prefetches that ignore cache reuse.
    
    \item We propose a RISC-V matrix ISA that densifies multiple sparse operations into a single dense operation, thereby improving PE utilization (Fig.~\ref{fig_moti}(c) upper).
    
    \item We design a lightweight runahead execution scheme for MPUs with an effective filter that improves area, performance, and energy efficiency (Fig.~\ref{fig_moti}(c) lower).
    
    \item Experimental results show that DARE improves performance between 1.04$\times$ and 4.44$\times$ and energy efficiency between 1.00$\times$ and 22.8$\times$ over the baseline, while reducing hardware overhead by 3.19$\times$ compared to NVR.
\end{itemize}

\section{Background and Motivation}

\subsection{Matrix ISAs in CPUs}

Recently, CPU vendors have integrated matrix ISA extensions, beyond traditional vector ISAs and processors~\cite{10992987,10992880,10546713,10546747}. We use Intel's AMX~\cite{b_amx} as a reference. It introduces eight 1~KB matrix registers, each with 16 rows and 64 bytes per row, and provides instructions for memory access and Matrix-Multiply-Accumulate (MMA), all executed at the matrix-register granularity. The memory access instructions require a uniform address stride across rows, which simplifies the ISA design.

\subsection{Irregularity in DNNs and ISA Constraints}\label{sec:isaConstrain}

Sparsity is widely adopted in DNNs to reduce computational cost~\cite{b_sparse1}, resulting in computation kernels with irregular memory access patterns and computation flows, exemplified by SpMM and SDDMM \cite{rahman2021fusedmm,nisa2018sampled}. An example computation flow of SDDMM\footnote{We omit SpMM here due to page limits. Its computation flow can be found in~\cite{b_spade}.} is shown in Fig.~\ref{fig_moti}(a), where computation is required only at the non-zero positions of matrix~C. The load of matrix~A involves two levels of indirection imposed by the Compressed Sparse Column (CSC) format\footnote{The compressed sparse row format also imposes indirection \cite{10546797}. We here take CSC as an exmaple.}, which is difficult to predict. In summary, sparsity poses significant challenges to PE utilization in systolic arrays and to memory access efficiency.

\textbf{Current Matrix ISAs would benefit from non-strided memory access instructions to implement densification.} Block-wise sparsity~\cite{b_block_sparse1,b_block_sparse2,yuan2025native} can improve utilization but may introduce redundant computation. We observe that multiple sparse MMA operations can be densified into a single MMA instruction to achieve higher PE utilization. For instance, the computation of the first column of matrix~C in Fig.~\ref{fig_moti}(a) can be densified as shown in the upper part of Fig.~\ref{fig_moti}(c). Our experiments show that PE utilization can improve by up to 14.4$\times$ when densification is fully exploited. However, unequal address strides between rows~0, 1, and~3 of matrix~A prevent densification under current ISA stride constraints, resulting in a two-step execution shown in the upper part of Fig.~\ref{fig_moti}(b). This motivates the need for ISA extensions that support non-strided memory access to improve PE utilization.

\subsection{Runahead Execution and Limitations}\label{sec:re}

Runahead execution~\cite{b_rh,b_pre,b_vr,b_dvr,b_svr} is a pre-execution prefetching technique for irregular memory accesses. It allows the processor to speculatively execute future instructions when stalled by long-latency memory accesses. The memory requests generated during this runahead phase provide accurate prefetches since they follow the actual control flow. NVR~\cite{b_nvr} extends this idea to DNN workloads by applying runahead on an NPU. However, porting such techniques to an MPU presents significant challenges.

\begin{figure}
    \centering
    \includegraphics[width=1.0\linewidth]{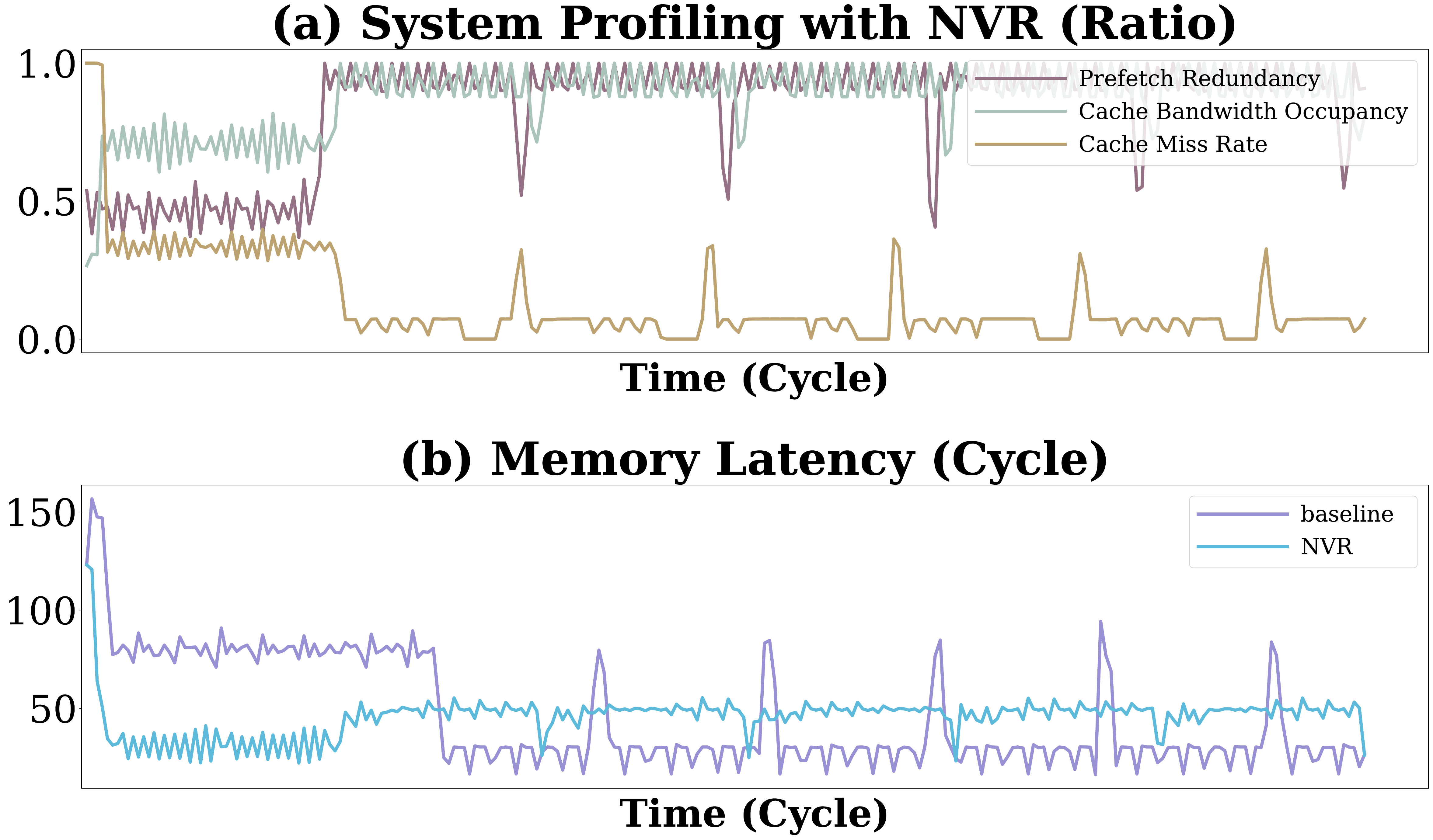}
    \caption{\textbf{(a)} {The cache miss rate, prefetch redundancy and the cache bandwidth occupancy in NVR on SDDMM.} \textbf{(b)} {The average memory access latency in baseline and NVR.}}
    \label{fig_bg_analyze}
\end{figure}

\textbf{Inefficiency from Prefetch Redundancy.} A redundant prefetch occurs when a prefetched block is already present in the cache. Fig.~\ref{fig_bg_analyze}(a) shows that runahead produces excessive redundant prefetches when the cache miss rate is low. They contend for cache bandwidth like normal requests and can eventually saturate it. As shown in Fig.~\ref{fig_bg_analyze}(b), this redundancy substantially increases memory latency and degrades performance. The root cause lies in data reuse within DNN workloads. The lower part of Fig.~\ref{fig_moti}(b) illustrates an example: although only column~1 of matrix~B misses in iteration~2, NVR prefetches both row~0 of matrix~A and column~1 of matrix~B. Row~0 of matrix~A was already prefetched in iteration~1 and remained in the cache, making the prefetch redundant. Processing units with smaller local memories (registers or scratchpads) suffer more redundancy because they rely on caches for reuse. Thus, redundancy becomes more severe when moving from a large NPU with a 256~KB scratchpad such as Gemmini~\cite{b_gemmini} to a smaller MPU with 8~KB registers such as AMX. Because data reuse is common in DNN workloads, redundancy broadly degrades both performance and energy efficiency.

\textbf{High Implementation Overhead.} Most prior runahead techniques rely on checkpointing because they speculatively pre-execute future instructions and require context restoration. Checkpoint register files can incur up to 8~KB overhead in an AMX-like design. Some approaches, such as NVR, avoid checkpointing but incur a higher overhead of 9.72~KB. Thus, existing runahead implementations remain too costly for MPUs.


The inefficiency in area, performance, and energy consumption of deploying runahead on MPUs calls for a more lightweight design with a runahead filter to fully exploit runahead execution.

\section{DARE Instruction Set Architecture} \label{sec_isa}

In this section, we first introduce the basic DARE ISA. Then we extend it to support non-strided memory access for densification.

\subsection{Basic Instruction Set Architecture}

The DARE ISA is a RISC-V matrix ISA inspired by Intel AMX. It provides eight 1~KB matrix registers (\textit{m0}-\textit{m7}), each organized as 16 rows by 64 bytes, and three Control and Status Registers (CSRs), \textit{matrixM}, \textit{matrixK}, and \textit{matrixN}, which define the logical shape of the matrix registers.

To support load, store, and MMA operations, DARE introduces three core instructions: \texttt{mld}, \texttt{mst}, and \texttt{mma}. \texttt{mld} loads a tile from memory into a matrix register, and \texttt{mst} stores a tile from a matrix register to memory. Each tile consists of \textit{matrixM} rows and \textit{matrixK} bytes per row. Both \texttt{mld} and \texttt{mst} take two source operands from general-purpose registers: one holds the base address of the first row and the other holds the stride between consecutive rows. \texttt{mma} performs an MMA across three matrix registers. The shape of two source registers is \textit{matrixM}$\times$\textit{matrixK} and \textit{matrixN}$\times$\textit{matrixK}, respectively. In addition, a configuration instruction, \texttt{mcfg}, is provided to configure the CSRs.

\subsection{Supporting Non-Strided Memory Access}

%

We extend the basic ISA to support non-strided memory access by treating the first element of each matrix register row as a base address vector. We refer to this extension as Gather Scatter Access (GSA) and define the corresponding non-strided load and store instructions as \texttt{mgather} and \texttt{mscatter}, respectively. \texttt{mgather} and \texttt{mscatter} use the elements of this base address vector as per-row base addresses for memory load and store operations. The base address vector itself can be loaded from memory using \texttt{mld}. The address generation overhead can be alleviated by decoupling address generation from MPU execution into two threads, while the hardware overhead requires only minor modifications to the decoder’s handling of source operands compared to \texttt{mld} and \texttt{mst}. The complete DARE ISA is summarized in Table~\ref{tab_isa}.


\begin{table}[htbp]
    \centering
    \caption{DARE Instructions List}
    \begin{tabular}{cc}
        \toprule
        \textbf{Assembly Format} & \textbf{Description}  \\
        \midrule
        mcfg, rs1, rs2 & Write the value in rs2 to the CSR indexed by rs1 \\
        mld, md, (rs1), rs2 & Load a tile from address rs1 with rs2 stride to md \\
        mst, ms3, (rs1), rs2 & Store a tile to address rs1 with rs2 stride from ms3 \\
        mma, md, ms1, ms2 & Multiply ms1 and ms2 and accumulate to md \\
        mgather, md, (ms1) & Load a tile addressed by ms1 to md \\
        mscatter, ms2, (ms1) & Store a tile addressed by ms1 from ms2 \\
        \bottomrule
    \end{tabular}
    \label{tab_isa}
\end{table}

%


\begin{figure*} [htbp]
    \centerline{\includegraphics[width=1.0\linewidth]{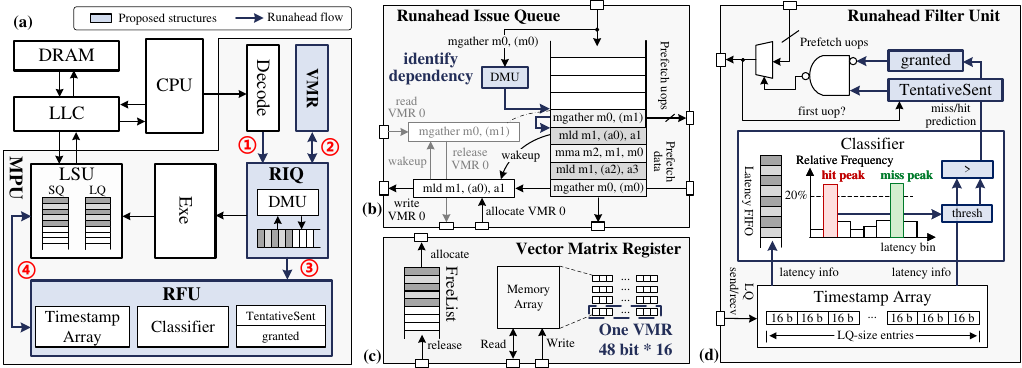}}
    \caption{\textbf{(a)} {Overview of the DARE architecture.} Blue blocks indicate components proposed by DARE. \textbf{(b)} {Runahead Issue Queue (RIQ) with a sub-module named the Dependency Management Unit (DMU).} \textbf{(c)} {Vector Matrix Register (VMR) as an auxiliary register file to store base vector addresses.} \textbf{(d)} {Runahead Filter Unit (RFU) with a threshold-based classifier.}}
    \label{fig_arch}
\end{figure*}

\section{DARE Microarchitecture} \label{sec_arch}

\subsection{DARE Overview}

An overview of the DARE microarchitecture is shown in Fig.~\ref{fig_arch}(a). Instructions are dispatched non-speculatively from the host CPU to the MPU. DARE is an out-of-order superscalar processor without register renaming. It connects directly to the Last Level Cache (LLC) through a Load Store Unit (LSU), which includes a Load Queue (LQ) and a Store Queue (SQ). All instructions are decomposed into micro-operations (uops) to simplify hardware implementation. Memory access instructions are further decomposed at the granularity of matrix register rows.

We introduce a circular queue named the \textbf{Runahead Issue Queue} (RIQ) that holds stalled instructions and serves as a candidate pool for prefetch uops. To enable runahead execution for \texttt{mgather}, we add a reduced matrix register file called the \textbf{Vector Matrix Register} (VMR), which provides temporary storage for base address vectors. To further mitigate performance and energy inefficiencies caused by prefetch redundancy, we extend the DARE architecture with a \textbf{Runahead Filter Unit} (RFU).

\subsection{Overall Workflow}

The CPU dispatches DARE instructions non-speculatively to the MPU, where each instruction is first decoded and then inserted into the RIQ~\circlednum{1}. Each memory access instruction in the RIQ is decomposed into prefetch uops, which are then arbitrated and filtered by the RFU~\circlednum{3}. Only uops that pass arbitration are issued to the LSU~\circlednum{4}. The head instruction of the RIQ is issued when it has no Read-After-Write (RAW), Write-After-Write (WAW), or Write-After-Read (WAR) conflicts with older instructions in the MPU. For \texttt{mgather} instructions in the RIQ, the RIQ identifies and wakes up the producer instruction that generates the base address vector. Each instruction in the dependency chain allocates a VMR entry~\circlednum{2} to hold its loaded data.



\subsection{Runahead Issue Queue}

The RIQ is a 32-entry circular queue, as determined in Section~\ref{sens_vmr_riq}, that accommodates stalled instructions together with a Dependency Management Unit (DMU), as shown in Fig.~\ref{fig_arch}(b). Each RIQ entry stores the full instruction information and a decompose counter. Memory instructions in the RIQ are decomposed into micro-operations (uops), which are sent to the RFU for arbitration.

When processing an \texttt{mgather} instruction, the DMU traverses the RIQ backward to identify the corresponding dependency chain, which terminates at an \texttt{mld}. It then wakes and executes the oldest instruction in the chain. Each completed instruction in turn wakes its consumer. Every woken instruction allocates a VMR entry and treats it as its destination register. A VMR entry is released once its consumer finishes reading the data.


\subsection{Vector Matrix Register}

The Vector Matrix Register (VMR) is a reduced matrix register file designed to enable accurate prefetching for \texttt{mgather} as it introduces matrix registers into the address dependency chain. Because the DARE architecture lacks additional physical registers, an auxiliary file is required to temporarily store base address vectors. Since only the first 48 bits of each row are needed under the Sv48 virtualization format with 48-bit virtual addresses, the VMR reduces the matrix register capacity per row to 48 bits, as shown in Fig.~\ref{fig_arch}(c). Each VMR entry is a 16-element vector corresponding to the 16 rows of a matrix register, with each element holding 48 bits. The VMR contains 16 entries in total, as determined in Section~\ref{sens_vmr_riq}. These entries are dynamically managed by a free list implemented as a circular queue that tracks the currently available VMR entries.

\begin{figure*}
\centerline{\includegraphics[width=1.0\linewidth]{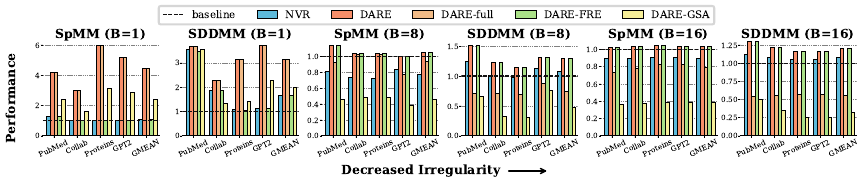}}
\caption{{Performance normalized to baseline.} DARE achieves 1.04$\times$ to 4.44$\times$ performance improvement on average compared to the baseline across various benchmarks. DARE is reported as the better between DARE-FRE and DARE-full. }
\label{fig_perf}
\end{figure*}

\subsection{Runahead Filter Unit}

The RFU, illustrated in Fig.~\ref{fig_arch}(d), filters redundant prefetch uops. It implements a \textbf{Tentative Uop Mechanism}: for each memory access instruction in the RIQ, only the first uop is issued initially, and subsequent uops are issued only if this tentative uop misses in the LLC or the instruction requires to write a VMR entry. Each RIQ entry is augmented with two fields: \textit{granted} and \textit{TentativeSent}. Uops are suppressed if the condition \textit{!granted~\&\&~TentativeSent} holds. The \textit{granted} flag is set when the tentative uop is classified as a LLC miss or when the instruction allocates a VMR entry, while the \textit{TentativeSent} flag is set when a corresponding uop is issued. The central challenge is \textbf{how to determine the miss/hit status of a uop, given that DARE cannot directly probe LLC status and memory conditions may vary at runtime}.

DARE addresses this challenge using a threshold-based, unsupervised binary classifier with uop latency as its only input. The key observation is that the memory access latency distribution typically exhibits a \textbf{bimodal shape}: one peak corresponds to LLC hits (red in Fig.~\ref{fig_arch}(d)) and the other to LLC misses (green in Fig.~\ref{fig_arch}(d)). The classifier dynamically updates a threshold such that any uop whose latency exceeds the threshold is classified as a cache miss. The threshold is updated in three steps:

\begin{enumerate}
  \item The classifier builds a histogram of recently observed latencies (32 in this paper) using bins of size 8 cycles.
  \item A bin whose relative frequency exceeds 20\% is identified as a peak. Only the smallest and largest peaks are retained.
  \item When the distance between the two peaks exceeds a margin (4 bins in this paper), the threshold is set to the latency of the minimum bin between them plus a fixed slack of 32 cycles. The slack prevents misclassifying a miss uop as a hit when hit latency fluctuates slightly.
\end{enumerate}

This dynamic classifier enables the RFU to adapt to runtime conditions, effectively reducing prefetch redundancy.

\section{Evaluation}

\subsection{Experimental Setup}

\subsubsection{Methodology}

We implement DARE in Register Transfer Level (RTL) and construct a cycle-accurate DARE model in gem5~\cite{b_gem5} to collect performance data. We synthesize DARE with Synopsys Design Compiler in a TSMC 28~nm process at 2~GHz to obtain power and area results. Cache timing and energy data are obtained from CACTI~7~\cite{b_cacti}. The detailed system configuration is listed in Table~\ref{tab_system_config}. We emulate NVR by equipping the MPU with infinite RIQ and VMR capacity, thereby preserving NVR's distant-prefetch capability, since baseline MPUs lack a specialized sparse unit. The baseline MPU excludes the RIQ, RFU, and VMR.


\begin{table}[t]
\caption{System configuration}
\centering
\begin{tabular}{l l}
\toprule
\textbf{Name} & \textbf{Detailed Configuration} \\
\midrule
Frequency & 2.0 GHz \\
\midrule
Host CPU & \makecell[l]{RV64GC + proposed DARE ISA, 8-way-issue out-\\of-order} \\
\midrule
MPU & \makecell[l]{48-entry LQ/SQ, 16$\times$16 systolic array with 32-bit-\\datapath PEs, 2-way-issue out-of-order} \\
\midrule
LLC & \makecell[l]{2MB, 16-way set associative, 16 banks, 1 read/1 \\write port, 20-cycle hit latency} \\
\midrule
Main Memory & \makecell[l]{45 ns latency, 50 GiB/s bandwidth} \\
\bottomrule
\end{tabular}
\label{tab_system_config}
\end{table}

For the ablation study, we define the following variants of DARE:

\begin{itemize}
    \item DARE-FRE: Only FRE is enabled.
    \item DARE-GSA: Only GSA is enabled.
    \item DARE-full: Both GSA and FRE are enabled.
    \item DARE: Better in DARE-FRE or DARE-full, as GSA can be disabled via an offline profiling.
\end{itemize}

\subsubsection{Benchmarks and Datasets}

We select two kernel-level benchmarks: SpMM and SDDMM. We choose graphs from PubMed \cite{b_pubmed}, OGBL-collab, and OGBN-proteins \cite{b_ogb}, and the attention map of GPT-2 \cite{b_gpt2} on Wikitext2 \cite{b_wiki} pruned to 90\% sparsity as the dataset. For the selected graphs, we take a subgraph from each to reduce simulation time. We further blockify the original datasets, with the notation $B=N$ indicating the block shape used to blockify is $N\times N$.

\subsection{Hardware Overhead}

The overall storage (Flip-Flop and SRAM) overhead is 3.05 KB, representing a 3.19$\times$ reduction compared to NVR. The total area overhead is 9.2\% relative to a baseline MPU, with 3.8\%, 4.1\%, and 1.3\% attributed to the VMR, RIQ, and RFU, respectively.


\subsection{Performance}

\subsubsection{Overall Performance} Performance results are shown in Fig.~\ref{fig_perf}. DARE achieves a geometric mean performance improvement of 1.04$\times$ to 4.44$\times$ across benchmarks compared to the baseline. DARE consistently outperforms both NVR and the baseline across all benchmarks. For specific benchmarks (e.g., SpMM with $B = 8$), NVR degrades performance (0.77$\times$) while DARE yields an improvement (1.05$\times$), underscoring DARE's effectiveness.

\subsubsection{Ablation Study} The ablation study results in Fig.~\ref{fig_perf} indicate that DARE-full performs best with highly irregular workloads, while DARE-FRE is more effective with lower irregularity. In highly irregular benchmarks ($B=1$), both DARE-FRE and DARE-GSA contribute to performance enhancement, and DARE-GSA outperforms DARE-FRE in such scenarios. Moreover, the performance enhancement of DARE-full (4.44$\times$) exceeds the product of DARE-FRE (1.06$\times$) and DARE-GSA (2.43$\times$) in SpMM, indicating a synergistic effect between them. When irregularity decreases to $B \ge 8$, the additional loading of base address vectors by DARE-GSA results in performance degradation, and DARE-FRE emerges as the dominant optimization. The ablation study results suggest that GSA should be selectively disabled based on workload characteristics, as will be discussed in Section~\ref{snes_block_size}.



\begin{figure}
    \centerline{\includegraphics[width=0.95\linewidth]{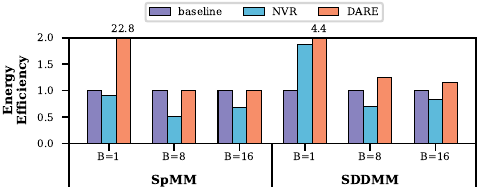}}
    \caption{{Energy efficiency normalized to the baseline.}}
    \label{fig_energy}
\end{figure}

\subsection{Energy Efficiency}

Fig.~\ref{fig_energy} depicts the energy efficiency results, showing that DARE achieves an average energy efficiency ranging from 1.00$\times$ to 22.8$\times$ compared to the baseline. In specific workloads such as SDDMM with $B=8$, NVR improves performance (1.09$\times$) at the cost of lower energy efficiency (0.71$\times$) due to its significant redundant prefetches. In contrast, DARE enhances both performance (1.29$\times$) and energy efficiency (1.25$\times$) by reducing this redundancy. Furthermore, DARE significantly enhances energy efficiency in SDDMM (4.37$\times$) and SpMM (22.8$\times$) with $B=1$ by reducing runtime and improving data reuse within the MPU. In benchmarks with low irregularity (SpMM with $B \ge 8$), DARE yields modest energy efficiency benefits due to the low cache miss rates in such scenarios.

\begin{figure}
    \centerline{\includegraphics[width=1.0\linewidth]{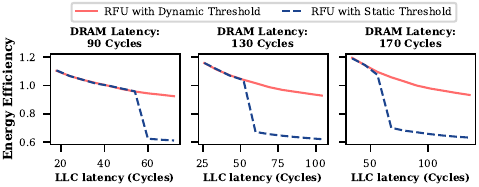}}
    \caption{{The robustness in energy efficiency of DARE across memory environments.}}
    \label{fig_robust}
\end{figure}

\subsection{Robustness across Memory Environments}

Fig.~\ref{fig_robust} illustrates the energy efficiency under various memory environments for SDDMM with $B=8$, normalized to the baseline. We implemented a baseline RFU with a static threshold of 64 cycles. In general, energy efficiency decreases with increased LLC latency, as the performance gain from FRE diminishes. However, the static-threshold RFU suffers an abrupt energy efficiency drop when LLC latency exceeds a value near its threshold, as it cannot distinguish between LLC hits and misses when LLC latency surpasses the static threshold, causing it to grant every instruction. In contrast, the dynamic-threshold RFU proposed in DARE demonstrates high robustness across various memory environments.

\subsection{Sensitivity to VMR Size and RIQ Size} \label{sens_vmr_riq}

\begin{figure}
\centerline{\includegraphics[width=1.0\linewidth]{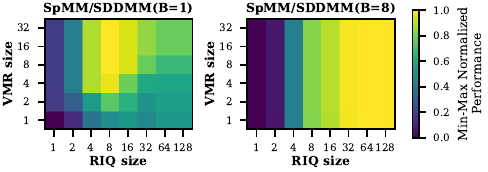}}
\caption{{The sensitivity of performance to VMR size and RIQ size.} Performance is normalized to [0,1] with the maximum and minimum. }
\label{fig_riq_vmr}
\end{figure}

Fig.~\ref{fig_riq_vmr} illustrates the sensitivity of DARE's performance to VMR and RIQ size, with performance normalized to the range [0,1] based on the maximum and minimum values for each case. The results suggest that the resource requirements for RIQ and VMR vary with the block size. When $B=1$, increasing the VMR size improves performance, while a larger RIQ may lead to performance degradation. This is because a larger RIQ increases the number of \texttt{mld} instructions targeting base address vectors, which are forced to be granted by the RFU. When $B=8$, performance is monotonic with respect to RIQ size, while VMR size has a negligible effect. This is because DARE-FRE dominates in this scenario, with no requirement for the VMR but a necessity for a large RIQ to hide the prediction latency of the RFU. To balance these two scenarios, we chose a 32-entry RIQ and a 16-entry VMR for our design.

\begin{figure}
\centerline{\includegraphics[width=0.95\linewidth]{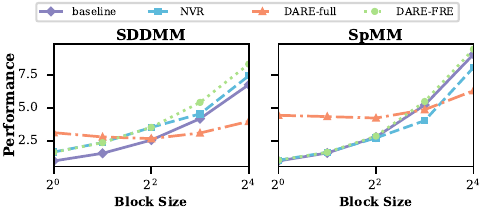}}
\caption{{The sensitivity of performance to the block size.} All results are normalized to baseline where $B=1$.}
\label{fig_block_sens}
\end{figure}

\subsection{Sensitivity to Block Size} \label{snes_block_size}

Fig.~\ref{fig_block_sens} shows the performance trends under different block sizes. The performance of DARE-full relative to DARE-FRE is monotonic with respect to block size, showing benefits at small $B$ and performance degradation at large $B$. Therefore, an offline profiling step can be used to decide when to disable GSA according to the block size: it is disabled at $B \ge 4$ for SDDMM and at $B \ge 8$ for SpMM. The results also show that using block-wise sparsity can effectively improve performance across workloads, as larger block sizes fit the systolic array better. DARE-FRE benefits from larger block sizes ($B \ge 8$) to a greater extent than both the baseline and NVR. Together, these results demonstrate DARE's adaptability to various block sizes.

%
%
%
%
%
\section{Conclusion}

This paper presents DARE, an irregularity-tolerant MPU that integrates a densifying ISA (GSA) and a filtered runahead mechanism (FRE) to improve PE utilization and reduce prefetch redundancy. Experimental results demonstrate that GSA and FRE act synergistically for highly irregular workloads ($B=1$), while FRE dominates when irregularity decreases ($B \ge 8$). Furthermore, DARE exhibits high robustness across various memory environments. Across SpMM and SDDMM benchmarks, DARE achieves up to a 4.44$\times$ performance improvement and a 22.8$\times$ energy efficiency enhancement over the baseline, with a 3.19$\times$ hardware overhead reduction compared to NVR. These results render DARE a practical solution for handling irregularity in DNN workloads.

\newpage

\bibliographystyle{IEEEtran}
\bibliography{refs}

\end{document}